\documentclass[trackchanges,twocolumn,tighten]{aastex63}

\newcommand{\ltsima}{$\; \buildrel < \over \sim \;$}
\newcommand{\simlt}{\lower.5ex\hbox{\ltsima}}

\def\arcdeg{\hbox{$^\circ$}}
\def\arcmin{\hbox{$^\prime$}}
\def\arcsec{\hbox{$^{\prime\prime}$}}

\usepackage{amsmath}
\usepackage{graphicx}

\turnoffedits{}

\begin{document}

\title{JWST Observations of Starbursts: PAHs Closely Trace the Cool Phase of M82's Galactic Wind}

\shorttitle{PAH Correlations in M82's Outflow}
\shortauthors{Lopez et al.}

\correspondingauthor{Sebastian Lopez}
\email{lopez.764@osu.edu}

\author[0000-0002-2644-0077]{Sebastian Lopez} 
\affil{Department of Astronomy, The Ohio State University, 140 W. 18th Ave., Columbus, OH 43210, USA}
\affil{Center for Cosmology and AstroParticle Physics, The Ohio State University, 191 W. Woodruff Ave., Columbus, OH 43210, USA}

\author[0009-0008-6851-4848]{Colton Ring} 
\affil{Department of Astronomy, The Ohio State University, 140 W. 18th Ave., Columbus, OH 43210, USA}

\author[0000-0002-2545-1700]{Adam K. Leroy}
\affil{Department of Astronomy, The Ohio State University, 140 W. 18th Ave., Columbus, OH 43210, USA}
\affil{Center for Cosmology and AstroParticle Physics, The Ohio State University, 191 W. Woodruff Ave., Columbus, OH 43210, USA}

\author[0000-0002-9511-1330]{Serena A. Cronin}
\affil{Department of Astronomy, University of Maryland, College Park, MD 20742, USA}

\author[0000-0002-5480-5686]{Alberto D. Bolatto} 
\affil{Department of Astronomy, University of Maryland, College Park, MD 20742, USA}
\affil{Joint Space-Science Institute, University of Maryland, College Park, MD 20742, USA}

\author[0000-0002-1790-3148]{Laura A. Lopez}
\affil{Department of Astronomy, The Ohio State University, 140 W. 18th Ave., Columbus, OH 43210, USA}
\affil{Center for Cosmology and AstroParticle Physics, The Ohio State University, 191 W. Woodruff Ave., Columbus, OH 43210, USA}

\author[0000-0002-5877-379X]{Vicente Villanueva}
\affil{Departamento de Astronomía, Universidad de Concepción, Barrio Universitario, Concepción, Chile}

\author[0000-0003-0645-5260]{Deanne B. Fisher}
\affil{Centre for Astrophysics and Supercomputing, Swinburne University of Technology, Hawthorn, VIC 3122, Australia}
\affil{ARC Centre of Excellence for All Sky Astrophysics in 3 Dimensions (ASTRO 3D)}

\author[0000-0003-2377-9574]{Todd A. Thompson}
\affil{Department of Astronomy, The Ohio State University, 140 W. 18th Ave., Columbus, OH 43210, USA}
\affil{Center for Cosmology and AstroParticle Physics, The Ohio State University, 191 W. Woodruff Ave., Columbus, OH 43210, USA}
\affil{Department of Physics, The Ohio State University, 191 W. Woodruff Ave., Columbus, OH 43210, USA}

\author[0009-0001-6065-0414]{Grant P. Donnelly}
\affiliation{Department of Physics \& Astronomy and Ritter Astrophysical Research Center, University of Toledo, Toledo, OH 43606, USA}

\author{Lee Armus}
\affil{IPAC, California Institute of Technology, 1200 E. California Blvd., Pasadena, CA 91125}

\author[0000-0002-5666-7782]{Torsten B\"oker}
\affil{European Space Agency, c/o STScI, 3700 San Martin Drive, Baltimore, MD 21218, USA}

\author[0000-0002-3952-8588]{Leindert A. Boogaard}
\affil{Max Planck Institut f\"ur Astronomie, K\"onigstuhl 17, D-69117 Heidelberg, Germany}

\author[0000-0003-4850-9589]{Martha L.\ Boyer}
\affiliation{Space Telescope Science Institute, 3700 San Martin Drive, 
Baltimore, MD 21218, USA}

\author[0000-0001-8241-7704]{Ryan Chown} 
\affil{Department of Astronomy, The Ohio State University, 140 W. 18th Ave., Columbus, OH 43210, USA}

\author[0000-0002-5782-9093]{Daniel A. Dale}
\affiliation{Department of Physics \& Astronomy, University of Wyoming, Laramie, WY 82071}

\author[0009-0004-5807-9142]{Keaton Donaghue}
\affiliation{Department of Astronomy, University of Maryland, College Park, MD 20742, USA}

\author[0000-0001-6527-6954]{Kimberly Emig}
\affil{National Radio Astronomy Observatory, 520 Edgemont Road, Charlottesville, VA 22903, USA}

\author{Simon C.O. Glover}
\affiliation{Universit\"{a}t Heidelberg, Zentrum f\"{u}r Astronomie, Institut f\"{u}r Theoretische Astrophysik, Albert-Ueberle-Str.\ 2, 69120 Heidelberg, Germany}

\author[0000-0002-2775-0595]{Rodrigo Herrera-Camus}
\affil{Departamento de Astronomía, Universidad de Concepción, Barrio Universitario, Concepción, Chile}
\affil{Millennium Nucleus for Galaxies (MINGAL), Concepción, Chile}

\author[0000-0002-0560-3172]{Ralf S.\ Klessen}
\affiliation{Universit\"{a}t Heidelberg, Zentrum f\"{u}r Astronomie, Institut f\"{u}r Theoretische Astrophysik, Albert-Ueberle-Str.\ 2, 69120 Heidelberg, Germany}
\affiliation{Universit\"{a}t Heidelberg, Interdisziplin\"{a}res Zentrum f\"{u}r Wissenschaftliches Rechnen, Im Neuenheimer Feld 225, 69120 Heidelberg, Germany}

\author[0000-0001-8490-6632]{Thomas S.-Y. Lai }
\affil{IPAC, California Institute of Technology, 1200 E. California Blvd., Pasadena, CA 91125}

\author[0000-0003-4023-8657]{Laura Lenki\'{c}}
\affil{IPAC, California Institute of Technology, 1200 E. California Blvd., Pasadena, CA 91125}

\author[0000-0003-2508-2586]{Rebecca C. Levy}
\affiliation{Space Telescope Science Institute, 3700 San Martin Drive, Baltimore, MD 21218, USA}

\author[0000-0001-9436-9471]{David S. Meier}
\affiliation{New Mexico Institute of Mining and Technology, 801 Leroy Place, Socorro, NM 87801, USA}

\author[0000-0001-8782-1992]{Elisabeth Mills}
\affiliation{Department of Physics and Astronomy, University of Kansas, 1251 Wescoe Hall Drive, Lawrence, KS 66045, USA}

\author[0000-0001-8224-1956]{Juergen Ott}
\affil{National Radio Astronomy Observatory, PO Box O, Lopezville Road 1011, Socorro, NM 87801,  USA}

\author[0000-0003-0605-8732]{Evan D. Skillman}
\affil{Minnesota Institute for Astrophysics, University of Minnesota, 116 Church St. SE, Minneapolis, MN 55455, USA}

\author[0000-0003-1545-5078]{J. D. T. Smith}
\affil{Department of Physics \& Astronomy and Ritter Astrophysical Research Center, University of Toledo, Toledo, OH 43606, USA}

\author[0000-0003-1356-1096]{Elizabeth J.\ Tarantino}
\affiliation{Space Telescope Science Institute, 3700 San Martin Drive, 
Baltimore, MD 21218, USA}

\author[0000-0002-3158-6820]{Sylvain Veilleux} 
\affil{Department of Astronomy, University of Maryland, College Park, MD 20742, USA}
\affil{Joint Space-Science Institute, University of Maryland, College Park, MD 20742, USA}

\author[0000-0003-4850-9589]{Fabian Walter}
\affiliation{Max Planck Institut f\"ur Astronomie, K\"onigstuhl 17, D-69117 Heidelberg, Germany}
\affiliation{California Institute of Technology, Pasadena, CA 91125, USA}

\author[0000-0003-4850-9589]{Paul P.~van der Werf}
\affil{Leiden Observatory, Leiden University, PO Box 9513, 2300~RA Leiden, The Netherlands}

\begin{abstract}

Stellar feedback drives multiphase gas outflows from starburst galaxies, but the interpretation of dust emission in these winds remains uncertain. To investigate this, we analyze new JWST mid-infrared images tracing polycyclic aromatic hydrocarbon (PAH) emission at 7.7 and 11.3~$\mu$m from the outflow of the prototypical starburst M82 out to $3.2$ kpc. We find that PAH emission shows significant correlations with CO, H$\alpha$, and X-ray emission within the outflow, though the strengths and behaviors of these correlations vary with gas phase and distance from the starburst. PAH emission correlates strongly with cold molecular gas, with PAH--CO scaling relations in the wind nearly identical to those in galaxy disks despite the very different conditions. The H$\alpha$--PAH correlation indicates that H$\alpha$ traces the surfaces of PAH-bearing clouds, consistent with arising from ionized layers produced by shocks. Meanwhile the PAH–X-ray correlation disappears once distance effects are controlled for past 2~kpc, suggesting that PAHs are decoupled from the hot gas and the global correlation merely reflects the large-scale structure of the outflow. The PAH-to-neutral gas ratio remains nearly flat to 2~kpc, with variations following changes in the radiation field. This implies that the product of PAH abundance and dust-to-gas ratio does not change significantly over the inner portion of the outflow. Together, these results demonstrate that PAHs robustly trace the cold phase of M82’s wind, surviving well beyond the starburst and providing a powerful, high-resolution proxy for mapping the life cycle of entrained cold material in galactic outflows.

\end{abstract}


\section{Introduction}

\begin{figure*}
    \centering
    \includegraphics[width=\textwidth]{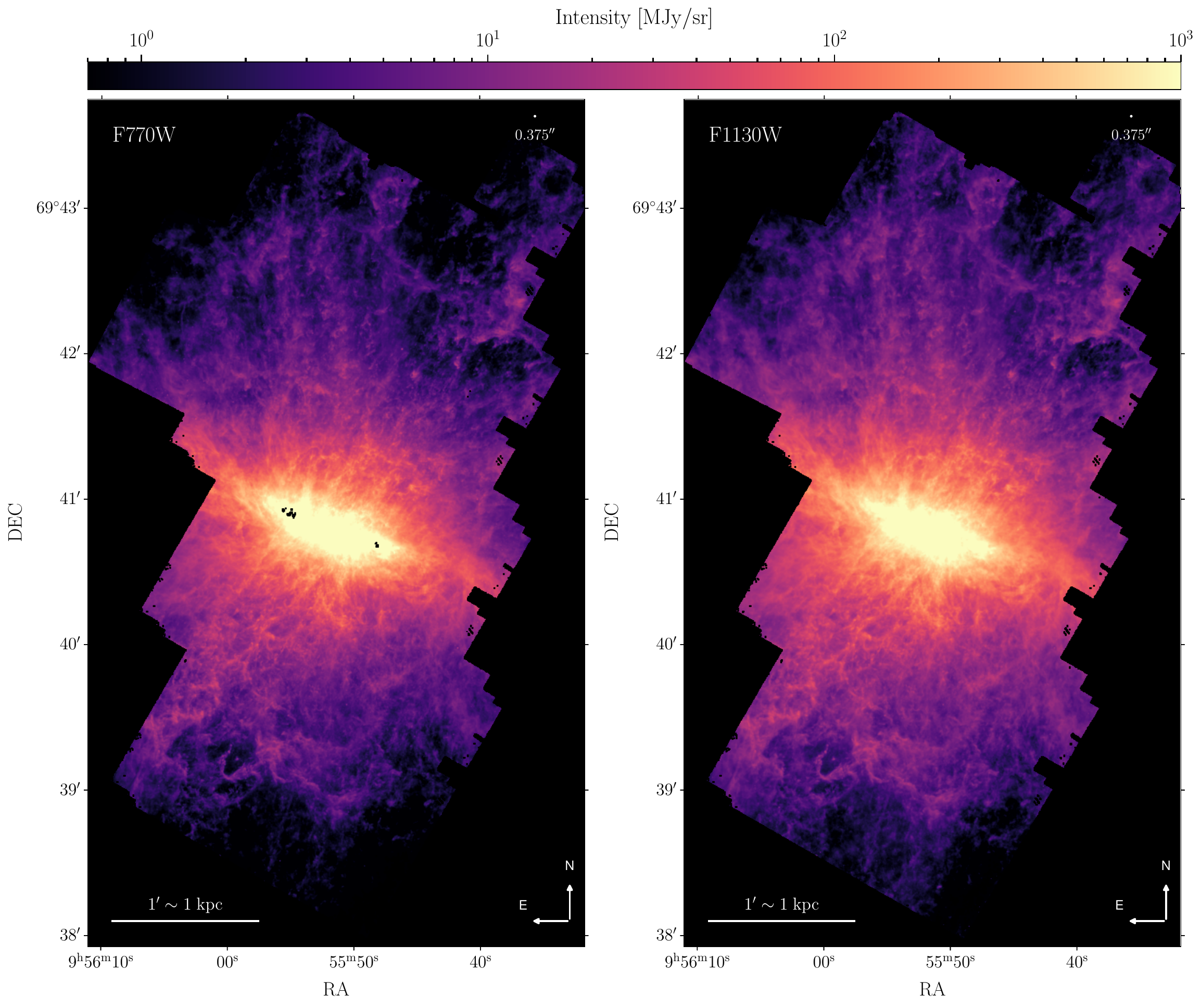}
    \caption{\textit{Left}: Log-stretched JWST F770W image of M82 with dust and stellar continuum and background subtracted, capturing the strength of the 7.7$\mu$m PAH feature. Blank pixels in the center are where the detector is saturated. \textit{Right}: Same as left but for the F1130W data tracing the 11.3$\mu$m PAH feature. In the top right the PSF of the final images is shown.}
    \label{fig:jwstdata}
\end{figure*}

\begin{figure*}[ht!]
    \centering
    \includegraphics[width=\textwidth]{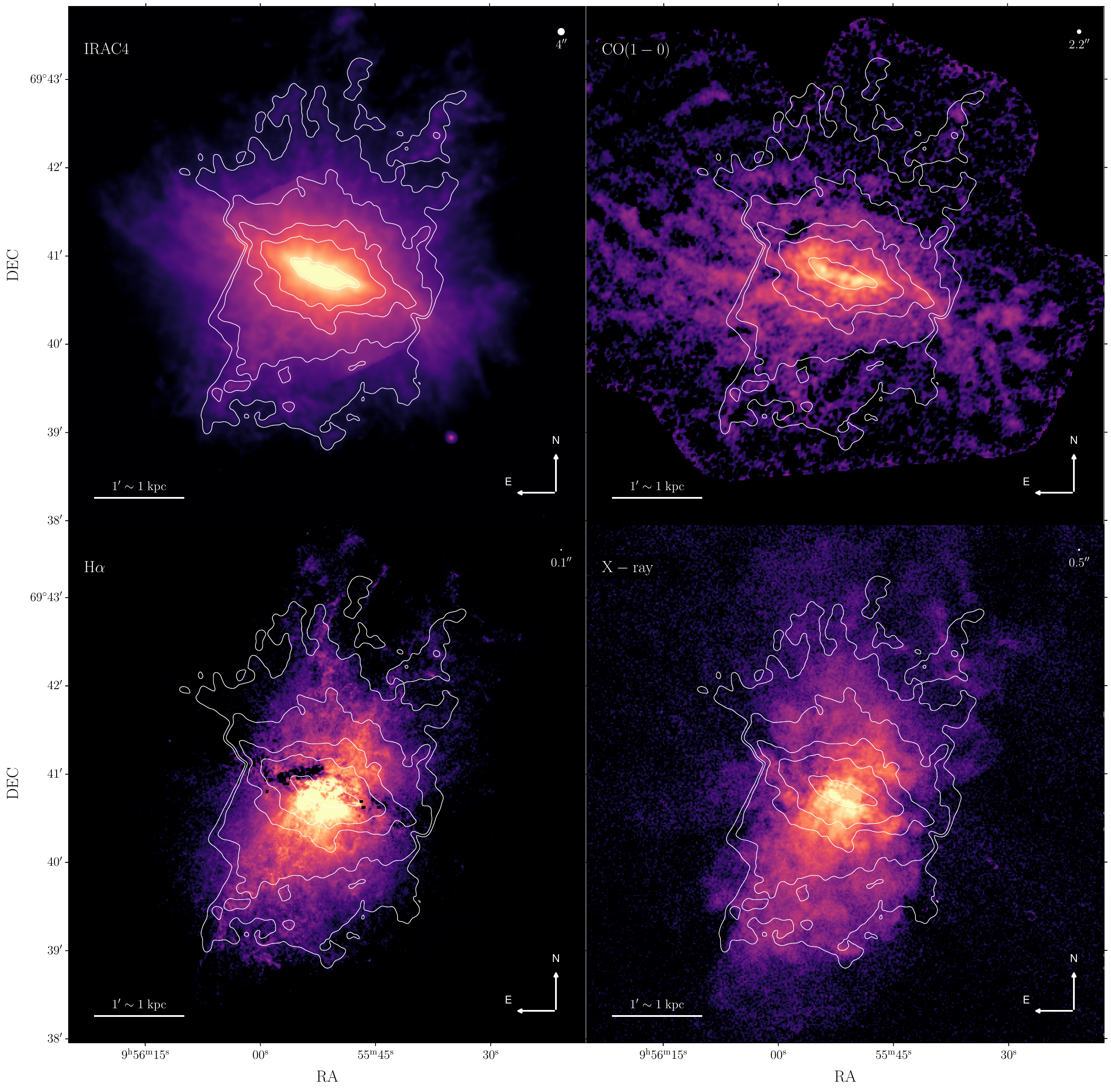}
    \caption{Log-stretch images of M82's multiphase outflow. \textit{Top Left}: \textit{Spitzer} 8~$\mu$m image from \cite{Engelbracht2006} reflecting mostly PAH 7.7~$\mu$m emission. \textit{Top Right}: IRAM NOEMA+30-m CO(1--0) emission from \cite{Krieger2021} tracing cold molecular gas. \textit{Bottom Left:} HST H$\alpha$ map tracing $\sim 10^4$~K ionized gas from \cite{Lopez2025}. \textit{Bottom Right}: \textit{Chandra} X-ray emission from \cite{Lopez2020} tracing hot $\sim 10^7$~K gas. The contours in all panels show the JWST 7.7~$\mu$m intensity (50, 75, 90, 95, 99 percentiles) as well as the field of view of the JWST data. The PSF size for each of images is shown in the top right.}
    \label{fig:data_image}
\end{figure*}

Galactic winds are a ubiquitous feature of starburst galaxies. They are driven by stellar feedback and play a vital role in redistributing metals, injecting energy and momentum into the circumgalactic medium (CGM), and regulating star formation \citep{Tumlinson2017, Veilleux2020}. These outflows are multiphase in nature, spanning large ranges of temperature, densities, and ionization states \citep{Heckman1990,Veilleux2005}. The hot $\sim 10^7$~K X-ray emitting phase, traces the supernova-heated, fast moving, gas that drives the wind \citep[e.g., see][]{Strickland2009,Lopez2020,Nguyen2021}. A warm $\sim 10^4$~K ionized gas phase is visible in optical line emission \citep[e.g.,][]{Xu2023,Lopez2025}, and can trace the interactions between the hot wind and colder material.

Cooler ($\sim 10^2{-}10^4$~K) atomic and molecular gas is seen in some outflows via 21-cm or molecular line emission \citep[e.g.,][]{Leroy2015,Martini2018,Krieger2021,Levy2023}. It may originate in the ISM \citep{Schneider2020} and be driven or entrained out of the galaxy, condense out of the hot wind \citep{Wang1995,Thompson2016}, or be present due to a combination of both effects. Over time, the cooler wind material may return to the galaxy akin to a galactic fountain, or it may be incorporated into the hot phase and reach the CGM. The large mass associated with this cold material compared to the hot phase makes understanding its life-cycle important \citep{Fluetsch2019,Herrera-Camus2020}.

Emission from dust grains in the infrared offers a promising direction to explore the physics of cold outflows because the grains are mixed with the outflowing gas \citep[e.g.,][in M82]{Engelbracht2006,Roussel2010}. With the launch of the James Webb Space Telescope (JWST), emission from polycyclic aromatic hydrocarbons (PAHs) has become a particularly powerful tool to study outflowing material at unprecedented angular resolution and sensitivity \citep{Bolatto2024,Fisher2025,Villanueva2025,Sutter2025}. These small grains produce spectral features at 3.3, 6.2, 7.7, 8.6, 11.2, 12.7, and 17.3~$\mu$m via their stretching and bending modes \citep[e.g.,][]{Puget1989,Tielens2008}. The illuminating radiation field, the abundance and ionization of PAHs, and dust-to-gas ratio also influence the intensity of PAH emission \citep[e.g., see][]{Leroy2023,Chown2025}.

In the prototypical starburst M82, previous high-resolution mid-infrared spectral mapping has already provided important insight into the behavior of PAHs in outflows. Using IRS integral-field spectroscopy of the central region, \citet{Beirao2008} showed that variations in PAH ratios can be driven by extinction and changes in the local radiation field. Subsequent IRS mapping of the full superwind revealed widespread PAH emission throughout the halo and strong spatial variations in grain size and ionization state \citep{Beirao2015}. These maps demonstrated that H$_2$/PAH ratios in extraplanar gas are enhanced, indicating that shocks contribute significantly to the excitation of the warm molecular component. The PAH 11.3/7.7 ratios were also found to increase with distance from the disk, pointing to larger and more ionized grains in the northern outflow.

In the star-forming disks of spiral galaxies, PAH emission correlates well with CO emission tracing molecular gas at 100~pc resolution \citep[e.g.,][and references therein]{Leroy2023,Sandstrom2023,Chown2025}. If a similar correlation holds in galactic outflows, JWST imaging can constrain the morphology, mass, and evolution of cold material as it moves away from the starburst without the need for expensive CO mapping. The fine-scale structure of cold gas is expected to bear the imprint of cloud launching, formation, and destruction, and of mass loading into the hot wind \citep[e.g., see][]{Thompson2024,Richie2025}.

In this paper, we measure how PAH emission and tracers of molecular, ionized, and hot gas correlate in M82's galactic outflow. JWST observations mapped PAH emission around M82 \citep{Bolatto2024, Villanueva2025} and revealed stunning filamentary structures \citep{Bolatto2024,Fisher2025} that often appear associated with molecular clouds \citep{Villanueva2025}. The IRAM telescopes imaged CO emission tracing the molecular gas across the same area \citep{Krieger2021}, HST mapped H$\alpha$ \citep{Mutchler2007,Lopez2025}, and \textit{Chandra} revealed the X-ray emission \citep{Lopez2020}. We combine these datasets to assess the relationship between PAHs and different gas phases in the starburst and outflow. We also compare the PAH-to-CO and PAH-to-gas ratios to radiation field estimates expected to alter the CO-PAH relationship. Throughout, we adopt a distance of 3.6~Mpc to M82, such that 1\arcmin\ $\approx$ 1~kpc \citep{Freedman1994}. We use the ICRS location as the center, $\mathrm{RA=09h55m52.43s}$ and $\mathrm{DEC=+69^{\circ}40\arcmin 46.93\arcsec}$ and midplane PA of $67.5^\circ$ \citep{Greco2012}.

\section{Data} \label{sec:data}

\begin{figure*}[ht!]
    \centering
    \includegraphics[width=0.88\textwidth]{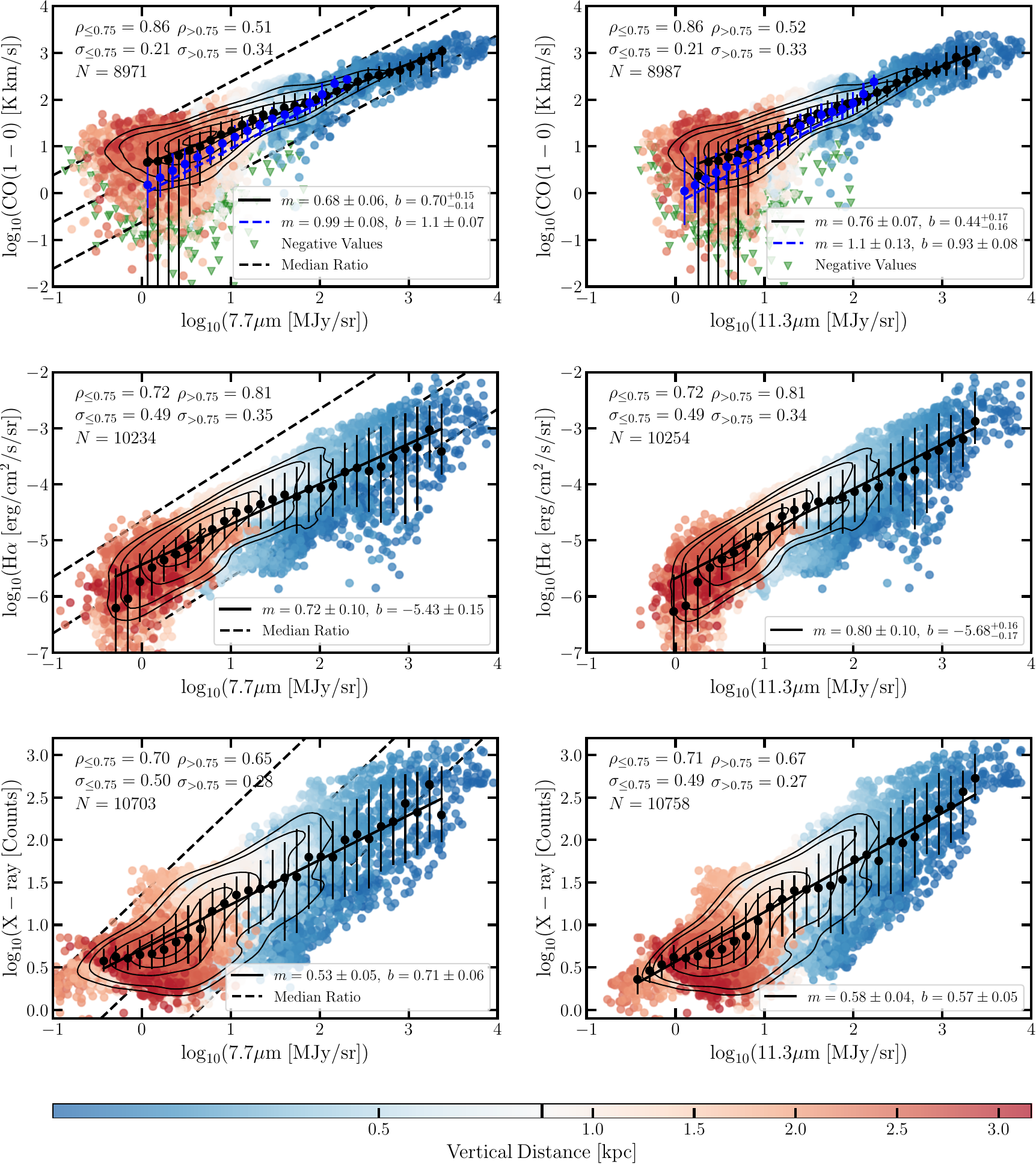}
    \caption{Correlations between PAH emission and gas tracers at $\sim38$~pc resolution. The \textit{left} column shows the relationship between JWST F770W and gas tracers, and the \textit{right} shows the same for F1130W. The color of points indicates their vertical distance from the M82 center. Overplotted are the median ratio of the bands along with the maximum and minimum of the colorbar in Figure~\ref{fig:ratios}. The \textit{top} row shows the relation between PAH and CO(1--0) emission, the \textit{middle} row compares to H$\alpha$, and the \textit{bottom} to X-ray emission. Black points in each panel show median $y$ values after binning the data by $x$, with error bars showing the 16{--}84\% range in each bin. The best-fit power law describing the bins is shown by the solid black line. Black contours enclose 15, 25, 50, 75, and 95\% of the data points, and green triangles in the top panels show the absolute value of CO data where the background-subtracted intensity is negative due to noise fluctuations. Dark blue points and dashed lines in the top row show the PAH-CO(2--1) relationships found for 70 galaxy disks by \citet{Chown2025}. In each panel we report Spearman rank coefficients ($\rho$), number of points in the correlation ($N$), and scatter ($\sigma$) of the data (in dex) about the best fit at distances above and below 750~pc (marked by the black line in the center of the colorbar). All gas tracers show good agreement with PAH emission, and the CO-PAH relationship closely resembles that found in galaxy disks. }
    \label{fig:whole_correlation}
\end{figure*}

\subsection{PAH Emission from JWST/MIRI} \label{ssec:jwst_miri}
We trace 7.7 and 11.3 um PAH emission using JWST images that cover the M82 outflow in the F770W and F1130W filters (doi:\dataset[10.17909/ew85-bc51]{https://doi.org/10.17909/ew85-bc51}) from \cite{Bolatto2024} (project GO 1701) which have previously appeared in \citet{Fisher2025} and \citet{Villanueva2025}. Following Cronin et al.\ (in preparation), we subtract a median background estimated from the outer part of the mosaic and confirm the estimate with the JWST Background Calculator\footnote{\url{https://github.com/spacetelescope/jwst_backgrounds}}. We verified that after this, the overall background level of the image appears consistent with previous wide-area \textit{Spitzer} imaging \citep{Engelbracht2006,Dale2009} to $\approx \pm 0.25$~MJy~sr$^{-1}$. We also account for the stellar and dust continuum that contribute to these PAH bands. Cronin et al.\ (in preparation) will provide the full details of the continuum subtraction, but in brief, we used the \textit{Spitzer} IRS spectral cubes of M82 with PAHFIT \citep{PAHFIT} to decompose each spectrum into its emission and continuum components. The resulting continuum maps were then regridded onto the JWST images, and a pixel-by-pixel continuum subtraction was applied.
The IRS-derived continuum maps overlap most of the JWST mosaic, but do not cover the full JWST field of view. Where IRS and JWST overlap, we apply the pixel-by-pixel continuum subtraction using the regridded IRS continuum maps. In regions outside the IRS footprint, we adopt a constant continuum fraction equal to the median measured within the IRS coverage for that morphological component. We find that the median continuum contribution to the F770W image is $\approx14\%$ in the northern outflow, $16\%$ in the disk, and $14\%$ in the southern outflow. For the F1130W image, the corresponding values are $\approx18\%$, $22\%$, and $18\%$. We define the disk as regions within $\pm750$~pc of the center of M82, with the remaining area corresponding to the northern and southern outflows.


\subsection{Ancillary High-Resolution Multiwavelength Observations} \label{ssec:ancillay_hires}

We compare PAH emission to CO(1--0) observations from \citet{Krieger2021}, H$\alpha$ emission from \cite{Lopez2025}, and X-rays from \cite{Lopez2020}. Full details of the data can be found in the corresponding references, and the following are brief summaries. We present the data products in Figure~\ref{fig:data_image}.

The CO(1--0) data include both IRAM NOEMA interferometric and 30-m single dish data, and we used the CASA \citep{CASA} task \texttt{feather} to combine the two datasets. Then we convolved the data to have a round synthesized beam (2.2\arcsec) and estimated the noise ($\sim0.1-0.2$~MJy/sr) based on signal-free regions of the data cube. We produced integrated intensity (moment 0) and associated uncertainty maps by integrating within a three-dimensional mask that includes all position-position-velocity locations where the HI 21-cm map from \citet{Martini2018}, the previous IRAM 30-m  CO(2--1) map \citep{Leroy2015}, or the \citet{Krieger2021} 30-m CO(1--0) show high significance emission at $27\arcsec$ resolution. This yields a high completeness map of CO(1--0) at $2.2\arcsec$ resolution (Figure~\ref{fig:data_image}). Hereafter we use CO(1--0) and CO interchangeably unless specified (i.e., CO(2--1)).

The H$\alpha$ image from \cite{Lopez2025} was observed by the Hubble Heritage Team and mosaicked by \cite{Mutchler2007}. \cite{Lopez2025} performed continuum subtraction, removed point sources, corrected for foreground extinction, and removed the [\ion{N}{2}] contamination using narrow-band data from the pathfinder-Dragonfly Spectral Line Mapper \citep{Lokhorst2022}. 

The Chandra X-ray data were presented by \cite{Lopez2020}, using nine archival observations totaling 534~ks of exposures and cover the $0.5-7$~keV band (doi:\dataset[10.25574/cdc.320]{https://doi.org/10.25574/cdc.320} ). The observations were astrometrically aligned, merged, exposure-corrected, and point sources were removed. We matched the JWST, HST H$\alpha$, and \textit{Chandra} images to the $2.2\arcsec$ resolution of the CO data following the procedures in \citet{Williams2024}. 

\subsection{Ancillary Low-Resolution Multiwavelength Observations} \label{ssec:ancillary_lowres}

We also compare the PAH emission to other lower-resolution multiwavelength observations as a function of distance along the outflow. PAH emission should be influenced by the radiation field, and so we consider maps of radiation field estimates from dust infrared SED modeling \citep{Leroy2015} and from pointed [C\,\textsc{ii}] 158\micron\ observations and photodissociation region (PDR) modeling \citep{Levy2023}. In addition, we compare to wide-area, lower-resolution \textit{Spitzer} 8 $\mu$m data that trace the 7.7 $\mu$m PAH feature \citep{Engelbracht2006,Dale2009} doi:\dataset[10.26131/IRSA424]{https://doi.org/10.26131/IRSA424}. Finally, to place the PAH emission in the context of the neutral gas reservoir, we use HI 21 cm maps from the VLA+GBT \citep{Martini2018} and CO(2--1) maps from IRAM \citep{Leroy2015}. 

Because the multiwavelength data have lower angular resolutions than JWST, we make versions of the data at common resolutions of $2.2\arcsec$ (including JWST, H$\alpha$, CO, and X-ray), $4\arcsec$ (including \textit{Spitzer}), and $27\arcsec$ (including all data). In each case we convolve the data to the coarser PSF and reproject the data onto a shared astrometric grid.

\section{Results and Discussion}

\subsection{PAH-Gas Phase Correlation}

\label{ssec:pah_correlations}

In Figure~\ref{fig:jwstdata} we present the JWST F770W and F1130W images, and Figure~\ref{fig:data_image} shows the \textit{Spitzer} $8\mu$m, CO(1--0), H$\alpha$, and X-ray images with the JWST 7.7~$\mu$m contours overlaid. 
The F770W, F1130W, and \textit{Spitzer} $8\mu$m filters all track PAH emission and each other well throughout the outflow. The JWST contours corresponds well to the CO in the southern outflow. The broader \textit{Spitzer} shows the good correspondence between the PAHs and the CO map in the extended eastern/western disk.
We also find a good match with our JWST data and the H$\alpha$ in the outflow, with both images showing filamentary features and arcs particularly in the south. However there is relatively less H$\alpha$ compared to PAH emission at the edges of our JWST images and is further exacerbated when comparing with the \textit{Spitzer} image. The X-ray emission, like the H$\alpha$, appear narrower than the \textit{Spitzer} image, though the X-rays appear smoother than the PAH or H$\alpha$ emission, without prominent filamentary structure. 

In Figure~\ref{fig:whole_correlation} we plot the correlations between PAH emission and CO, H$\alpha$, and X-ray intensity. Because the JWST maps have much higher signal-to-noise than the other bands, we treat them as our independent ($x$) variable. To assess the average relationship for each correlation, we bin the data by PAH intensity and calculate the statistics for the $y$-axis quantity in each bin. For the CO(1--0) data, we also include negative values in the binning. These reflect the high noise in the CO map and must be included for the average to yield the correct median. Black points show the resulting median trends and the 16 to 84 percentile range. Then, we fit a line (in log-space) to the median-binned values with the form \begin{equation}
    \log_{10} ({I_{\rm comp}}) = m \log_{10} ({I_{\rm IR}\;[{\rm MJy/sr}]})+ b,
\end{equation}
where $I_{\rm IR}$ is the intensity of the PAH-tracing bands, $I_{\rm comp}$ is the intensity of the other bands being compared, $m$ is the slope, and $b$ is the intercept. For each panel we note the Spearman rank correlation coefficient ($\rho$) and the standard deviation ($\sigma$) about the best-fit relation estimated from the median absolute deviation at distances below and greater than 750~pc. We show the median ratio of the bands as dotted black lines and one dex above and below it. These ratios map to Figure~\ref{fig:ratios} for direct comparison.  

Our correlation analysis parallels that of \cite{Chown2025}, who compared PAH emission to CO(2--1) in $70$ galaxy disks. We compare to their results for the disks of galaxies, i.e. omitting galaxy centers. Their best-fit relation, shown in dark blue in the top row of Figure~\ref{fig:whole_correlation}, closely resembles what we find for the region of M82 covered by JWST in black. The overlap is excellent near the M82 disk (red points). At larger vertical heights (light and dark red), especially in the F770W filter, the M82 outflow is modestly brighter in CO at fixed PAH intensity than galaxy disks. This may be largely explained by a changing CO(2--1)-to-(1--0) ratio ($R_{21}$) in the wind\footnote{The figure adopts a constant $R_{21} = 0.65$ \citep[e.g.,][]{Leroy2022}, which is a reasonable first-order approximation for the \citet{Chown2025} sample. Future analysis should account for the variable $R_{21}$ in both their sample and in the M82 wind.}. 

We also consider a second possibility for the enhanced CO at fixed PAH intensity, which is PAH destruction in the outflow. \citet{Beirao2015} reported elevated H$_2$/PAH ratios in M82’s wind from \textit{Spitzer} IRS spectra, which could be interpreted as evidence for shocks that both enhance H$_2$ emission and destroy PAHs. The characteristic shock velocities inferred by \citet{Beirao2015} are relatively low ($v_{\rm shock}\sim40$~km~s$^{-1}$). This is below the regime where rapid, efficient PAH destruction is expected. However, \citet{Micelotta2010a} show that PAHs can still undergo non-negligible structural damage and erosion for shocks of order $v_{\rm shock}\sim50$~km~s$^{-1}$, with increasingly severe destruction at higher velocities. Therefore, while we do not expect shock-driven PAH destruction to be the dominant driver of the CO enhancement relative to the galaxy-disk relation, the \citet{Beirao2015} results imply that modest PAH processing in the outflow may contribute at some level to the observed trends.


We also quantify the scatter near and away from the starburst (above/below 750~pc) to test the robustness of the PAH-CO relationship. 
We adopt 750~pc as the boundary between the disk and the outflow for three reasons: first, the H\textsc{I} map used in our analysis goes into absorption within the inner $\sim$500~pc, leaving the central region undefined (see Fig.~\ref{fig:profiles}). Second, when we later convolve the data to a coarser $27\arcsec$ resolution for the vertical profiles, the resulting PSF corresponds to roughly 500~pc, and choosing a larger cutoff avoids contamination from the bright central disk. Finally, visual inspection shows that beyond $\sim750$~pc the disk no longer dominates the emission, and the observed signal is primarily from the outflow.
As shown in Figure~\ref{fig:whole_correlation}, the scatter increases away from the starburst. This widening envelope is visually apparent in the red points at large vertical distances in Figure~\ref{fig:whole_correlation}, which reflect both the lower signal-to-noise in the CO data (with noise-dominated measurements shown by the green triangles) and a real increase in intrinsic scatter with height. Assuming a Gaussian distribution, we run a Monte Carlo calculation adopting the power-law fit of CO vs.\ F770W as the true relation and adding the appropriate statistical noise to each measurement. We find that the scatter is mostly intrinsic\footnote{We calculate the intrinsic scatter as $\sigma_{\rm int} \approx \sqrt{\sigma^2_{\rm Total}-\sigma^2_{\rm CO}}$. We find $\sigma_{\rm int} \approx 0.21$ and 0.27 dex for the F770W-CO comparison at low and high elevations, respectively.} with only moderate statistical scatter increase with height ($\sim0.07$). The results are nearly identical for F1130W.

Figure~\ref{fig:whole_correlation} shows that within the region surveyed by JWST, PAH emission correlates well with H$\alpha$ and X-ray emission. Both show an approximate power-law relationship with the PAH emission. In contrast to CO, these bands both show more scatter at low elevation (red points), though the large range of intensities near the starburst mean that the correlation coefficients relating H$\alpha$ and X-rays to PAH emission are still high. These low elevation points include areas where PAH emission becomes much brighter relative to H$\alpha$ (low $\rm{I_{H\alpha}}$ and high $\rm{I_{PAH}}$). This likely reflects the impact of significant extinction on the H$\alpha$ while mid-IR PAH emission is comparatively less affected.

While a correlation between CO and PAH emission is expected at these spatial resolutions \citep{Bolatto2024,Fisher2025,Villanueva2025}, such close quantitative agreement between our results and \citet{Chown2025} is surprising. The PAH intensity should depend on the strength and hardness of the illuminating radiation field, the dust-to-gas ratio, and the abundance and physical state of the PAHs. These parameters might all be expected to change under the wide range of conditions found in the M82 outflow, as evidenced by the spatial variations in PAH band ratios reported by \cite{Beirao2015}, and to differ between the disks of galaxies and the M82 starburst and outflow. We explore the radiation field dependence in Section~\ref{ssec:scalings}.

The PAH-X-ray correlation is puzzling given that small dust grains like PAHs are not expected to survive exposure to the hot wind, as their sputtering times can be one to two orders-of-magnitude shorter than the dynamical time \citep{Richie2025}. The correlation may reflect that a large fraction of the soft X-ray emission is produced by hot gas interacting with cooler material via charge-exchange emission. Such emission has been found to be common in studies of X-ray emitting winds \citep{Okon2024}. In this case, while PAHs cannot survive on their own in the hot gas, if they are mixed with cooler material, their survival times are extended and interactions with the hot gas contribute to the charge-exchange emission. 

The H$\alpha$ emission is associated with cooler ionized gas and likely represents an intermediate case between CO and X-ray emission. As discussed by \citet{Lopez2025} and \citet[][for Pa$\alpha$]{Fisher2025}, much of the H$\alpha$ emission appears to arise from the surfaces of cold gas clouds that are ionized by shocks in the outflow \citep{Shopbell1998,Beirao2015}, though some may instead be photoionized closer to the starburst. The PAH and H$\alpha$ emission also display strikingly similar morphologies (Figure~\ref{fig:data_image}), suggesting a common origin. In Appendix~\ref{app:ha_zoomin}, we present a detailed view of one such structure (cloud 12; \citealt{Lopez2025}), where the H$\alpha$ emission lies along the wind-facing surfaces of the cold cloud traced by PAHs. This configuration highlights where the hot wind may be shocking colder material, and relatedly, producing the observed charge-exchange emission in the soft X-rays. Together, these results support the interpretation that both PAH and H$\alpha$ emission originate from the outer layers of cold clouds entrained in the hot outflow or located along the edges of the outflow cone.

We note that the coloring of points in Figure \ref{fig:whole_correlation} and the images in Figure \ref{fig:data_image} show that all bands decrease in intensity with increasing distance from the burst along the outflow. Therefore it is possible the correlations reflect a shared dependence of PAH and X-ray intensity on distance from the starburst, not any physical relationship between PAH emission and hot gas. Projection effects may contribute to this effect, as the typical biconical model for M82's outflow has a volume-filling hot phase with the cooler phases at the cone surface that would appear to overlap \citep[e.g., ][]{Leroy2015,Nguyen2022}. 

We attempt to control for this effect in Appendix~\ref{sec:append_distance}, where we examine the band-to-band correlations after accounting for the overall relationship of intensity with distance. We find only the CO and H$\alpha$ emission correlate past 2~kpc, whereas the X-ray becomes uncorrelated, indicating the decoupling of the PAHs and hot gas. These correlations at fixed elevation match the impression from previous multiwavelength comparisons targeting the outflow. Similar morphologies are apparent between the JWST PAH and HST H$\alpha$ images (Figure~\ref{fig:data_image}), particularly in the filaments and arcs of warm ionized gas highlighted by \cite{Bolatto2024}, \cite{Fisher2025}, and \cite{Lopez2025}. In particular, \cite{Fisher2025} found that the 3.3~\micron\ PAH feature correlates with the H$\alpha$ and CO emission at fixed distance but not with X-ray, consistent with our results. 

\subsection{Where are the PAHs?}
\label{ssec:ratios}

\begin{figure*}[ht!]
    \centering
    \includegraphics[width=\textwidth]{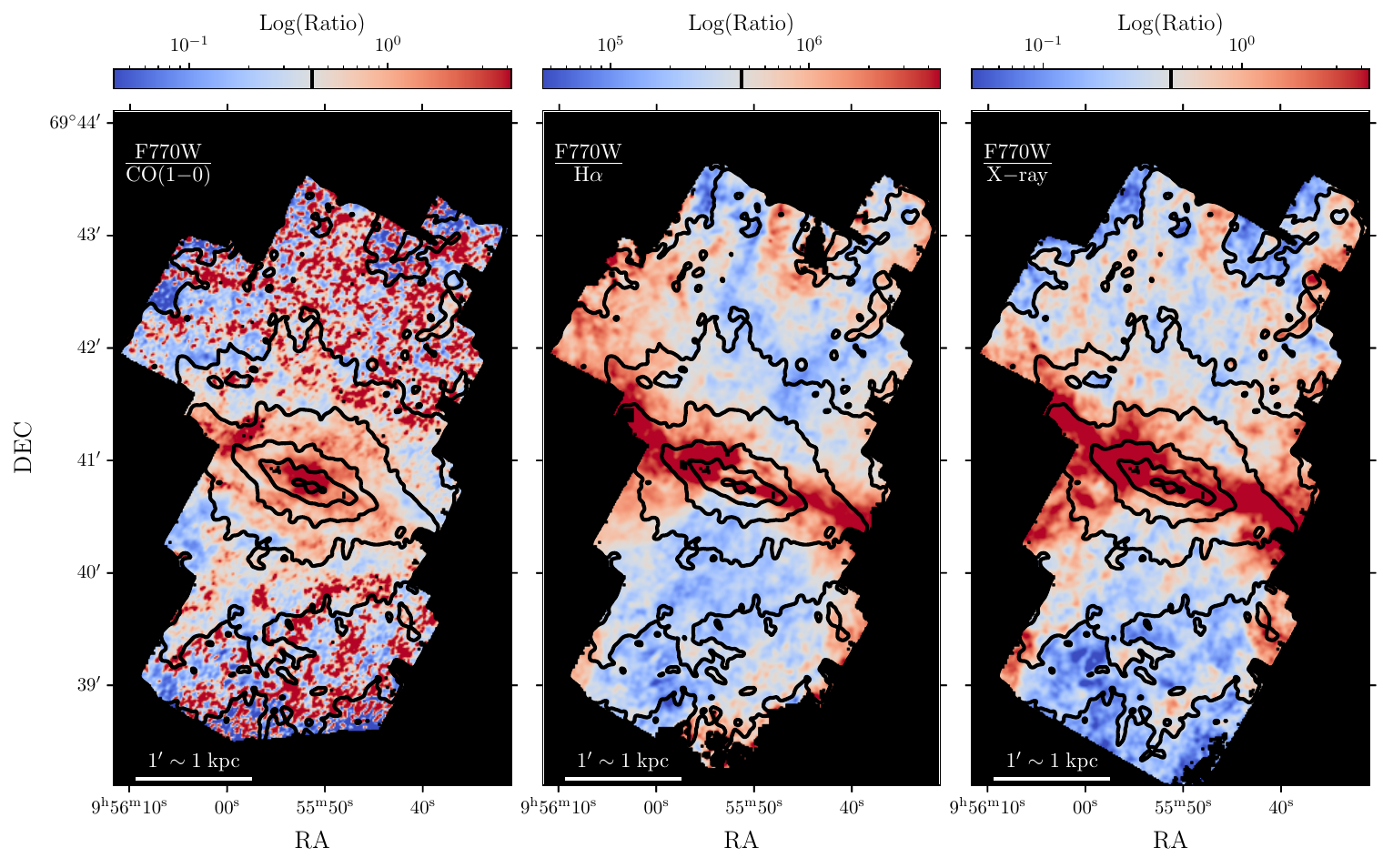}
    \caption{Ratio maps of JWST F770W to other wavelengths. All images were convolved to 2.2\arcsec{} and reprojected onto the same grid, and the ratio is shown on a log stretch centered on the median value (light blue), and the colorbar bounds are one dex below and above the median. Black contours mark the JWST F770W intensity. Within the central starburst, PAH emission appears enhanced relative to all other tracers. For the PAH-to-CO ($\mathrm{MJy/sr\;/\;K\;km/s}$), the red patches are where the CO emission is negative; thus they are set to upper limits. Elsewhere the CO and PAHs show a fairly constant ratio (light blue). Both PAH-to-H$\alpha$ ($\mathrm{MJy/sr\;/\;erg/cm^2/s/sr}$) and PAH-to-X-ray ($\mathrm{MJy/sr\;/\;Counts}$) ratios appear enhanced in the galaxy disk (due to extinction) and along the image edges (red), with patchy ratios (white and salmon) visible in the interior. This appears consistent with a picture where the PAH emission arises from cool material surrounding the hot wind or patchy, filamentary cool material is mixed into the wind.}
    \label{fig:ratios}
\end{figure*}

To better resolve the relative distribution of these tracers, Figure~\ref{fig:ratios} presents maps of the ratios between 7.7~$\mu$m~PAH emission and CO(1--0), H$\alpha$, and X-ray emission. These ratios were constructed by convolving to the CO resolution of 2.2\arcsec{}, reprojecting onto the CO map grid, and dividing the intensity maps. The color map is centered on the median ratio for each image, and the bounds are one dex below and above the median (marked in Figure~\ref{fig:whole_correlation} by the dashed black lines). 

The PAH-to-CO ratio map shows enhanced PAH emission in and around the starburst, then a large area of nearly fixed ratio (white and light blue or light red) around the disk. At larger distances, Figure \ref{fig:whole_correlation} shows that the median ratio remains constant, but the high noise in the CO map leads to many small patches with individually high (dark red) or low (dark blue) PAH-to-CO ratios. These patches reflect low signal-to-noise regions in the CO map, where negative CO intensities have been set to upper limits, and so should be interpreted as noise rather than real substructure. The ratio map does not show any strong pattern beyond this noise and the bright PAH emission in the burst, suggesting that PAHs and cold gas may remain well-mixed over the field. Such agreement is also found in the Makani Galaxy \citep{Veilleux2025} and demonstrates how PAH observations with JWST can provide a high-resolution proxy for cold gas observations.

The middle and right panels also show a complex picture for X-ray and H$\alpha$ emission. In both cases, PAHs are bright relative to the other band (red) in the starburst and the galaxy disk and are moderately brighter (salmon) along the edges of the outflow cone, while the other tracers are brighter inside the wind itself. The enhanced PAH emission compared to the ionized gas phases in the disk likely arises from a combination of effects. As shown by the H$\alpha$ image in Figure~\ref{fig:data_image}, prominent dust lanes are present in the disk of M82. This extinction suppresses the observed H$\alpha$ emission. A similar effect can attenuate the softer X-ray photons ($\lesssim2$~keV) that contribute significantly to the broad-band \textit{Chandra} image, leading to an apparent enhancement of PAH emission relative to the ionized gas in the disk.

The PAH/[\ion{Ne}{2}] ratio is similar to PAH/H$\alpha$ (\ion{Ne}{1} ionization potential is 21.6 ev) but less affected by extinction due to \ion{Ne}{2}'s wavelength. This ratio has been measured by \citet{Beirao2008} and \citet{Beirao2015} for M82’s disk and outflow, respectively. A comparison of these studies shows that PAH/[\ion{Ne}{2}] decreases by approximately an order of magnitude from the disk (PAH/[\ion{Ne}{2}]$\sim50$) to the outflow (PAH/[\ion{Ne}{2}]$\sim7$). This indicates a real physical change beyond only extinction. The higher ratio in the dist could reflect large amounts of diffuse PAH emission along the line of sight through the edge-on disk. Alternatively, PAH heating conditions may differ between the disk and the outflow, with a more intense and harder radiation field in the disk. Finally, the absorption of ionizing photons in dusty \ion{H}{2} regions in the dense disk may suppress the [\ion{Ne}{2}] emission.

It is also possible that PAH processing occurs within the outflow. Although the shock velocities inferred in M82's wind by \citet{Beirao2015} are relatively low, such shocks can still produce non-negligible processing of PAHs \citep{Micelotta2010a}, which may modestly reduce their emission. In addition, the radiation field decreases with increasing distance from the disk, and in Section~\ref{ssec:scalings} we show that this decline closely tracks the decrease in the gas-weighted PAH emission. These changes in the physical environment and in PAH properties can therefore contribute to the observed variations in the enhanced PAH-to-gas ratio maps. \edit1{Similar PAH behavior has been found in AGN-driven outflows, where regions affected by AGN feedback show deficits of ionized PAH features and altered PAH band ratios relative to star-forming regions \citep[e.g.,][]{Garcia2024,Zhang2024}. In those systems, radiative and shock-driven feedback modifies the PAH population, preferentially suppressing the ionized PAH carriers, indicating that the PAH properties observed in M82's wind may reflect a broader response of PAHs to energetic outflow environments.} A full analysis of the composition of PAHs in M82's wind will be presented in Cronin et al.\ (in preparation).


The enhanced PAH emission observed outside the outflow cone is consistent with the expected multiphase structure of galactic winds, in which the core is hot and diffuse while the outskirts are cooler and denser \citep{Thompson2024}.
Interestingly, enhanced PAH emission is also located within the outflow cone, appearing as patchy filaments. As mentioned in Section~\ref{ssec:pah_correlations}, these locations could be where PAHs are mixed with cooler gas where clouds are entrained in the hot outflow. These patches are less common in the PAH–H$\alpha$ ratio map because the two bands exhibit nearly identical morphologies, as shown in Figures~\ref{fig:jwstdata} and \ref{fig:data_image} and illustrated in Appendix~\ref{app:ha_zoomin} for an individual cloud.


\subsection{PAH Intensity, Gas Mass, and the Radiation Field}\label{ssec:scalings}
\begin{figure*}
    \centering
    \includegraphics[width=\textwidth]{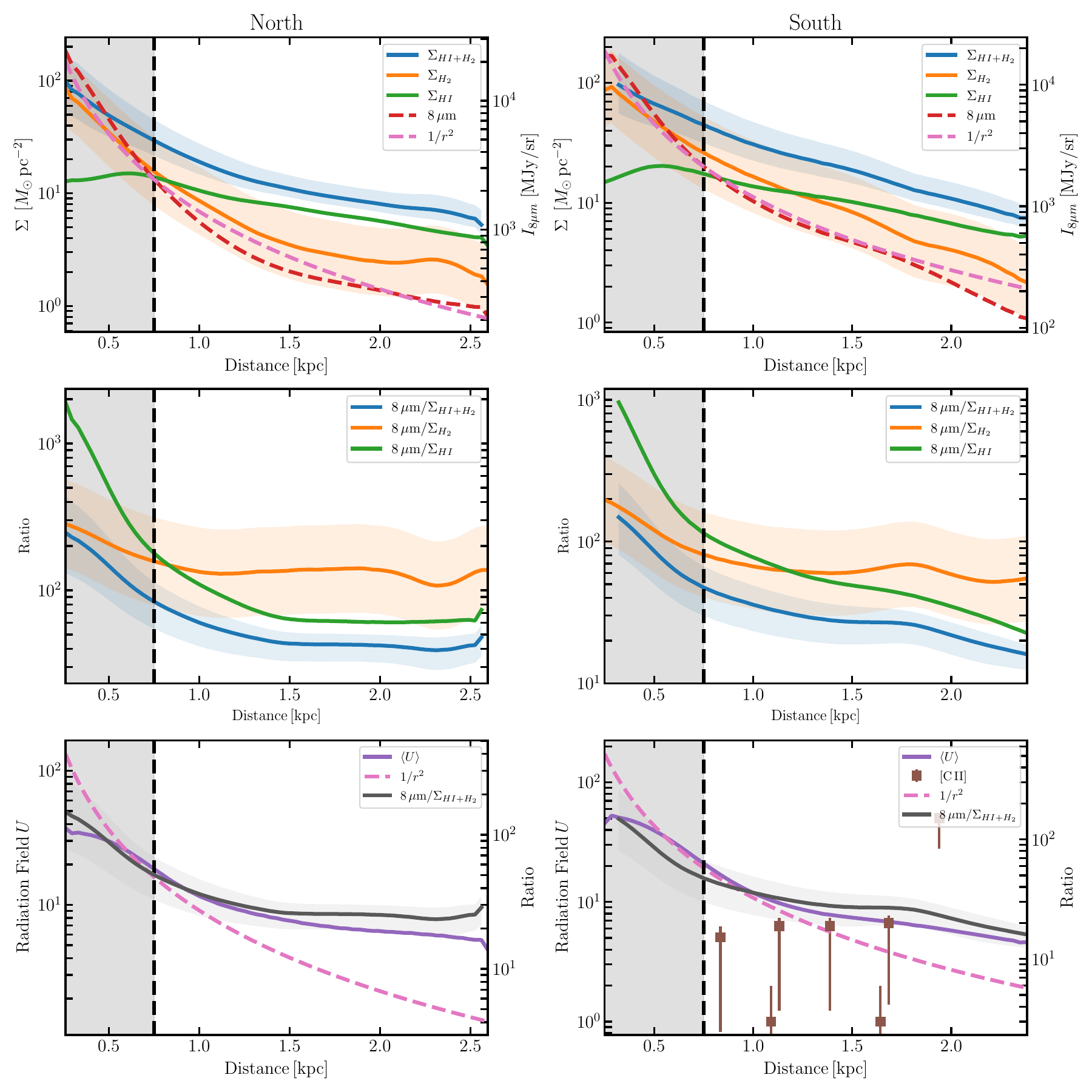}
    \caption{Profiles as a function of distance from the disk toward the north (left) and south (right). These profiles were constructed at a resolution of 27\arcsec{} ($\approx 500$~pc) to match the lowest resolution datasets. The profiles span the area covered by the JWST mapping (3.2\arcmin{}$\times$2.2\arcmin{} in the north and 2.4\arcmin{}$\times$2.0\arcmin{} in the south). We calculate the total gas surface density of the neutral phase as $\rm{\Sigma_{HI+H_2}}=\Sigma_{HI}+I_{CO(2-1)}\alpha_{CO}^{(2-1)}$ assuming $\rm{\alpha_{CO}^{(2-1)}=1\;M_\odot pc^{-2}}(K\;km\;s^{-1})^{-1}$ and an $R_{21}=0.6$ we derived from the IRAM CO(2--1) and CO(1--0) maps from \cite{Leroy2015} . For reference we add a vertical line at 750~pc since below that distance the HI goes into absorption and becomes unreliable. The \textit{top} row shows the mass surface densities of the HI, H$_2$, and the total neutral gas. Also shown is the fall off in intensity for the 8$\mu$m PAH emission from \textit{Spitzer}. The shaded areas around the H$_2$ and total gas mass are to mark the uncertainty in the $\alpha_{CO}$ used ($0.5<\alpha_{CO}<2$). The \textit{middle} row shows ratios that trace the coupling between PAH emission and the cold gas phases: 8~$\mu$m/H$_2$, 8~$\mu$m/HI, and 8~$\mu$m/${\mathrm{M_{HI+H_2}}}$. The \textit{bottom} row show different radiation field estimates: an analytic $1/r^2$ decline (dashed pink), $\langle U \rangle$ from dust SED modeling (solid violet), and $\mathrm{G/G_0}$ from [C II] PDR modeling (brown squares). The radiation field and PAH/$M_{\mathrm{HI+H_2}}$ ratio behave similarly, showing how the radiation field affects the PAH intensity.}
    \label{fig:profiles}    
\end{figure*}

The analyses in Sections~\ref{ssec:pah_correlations} and \ref{ssec:ratios} show that PAH emission correlates strongly with the CO(1--0) and H$\alpha$ emission near the disk and moderately so in the outflow. This suggests that PAHs are physically associated with the cool gas of the wind. However, the nearly fixed PAH-to-CO ratio seen in Figure~\ref{fig:whole_correlation} and \ref{fig:ratios} is unexpected. For stochastically heated grains, the intensity of PAH emission is expected to depend on local physical conditions as
\begin{equation}
\label{eq:ipah}
I_{\rm PAH} \propto U \times N(H) \times DGR \times q_{\rm PAH}\;,
\end{equation}
where $U$ is the intensity of the interstellar radiation field, $N(H)$ the column density of dust-bearing gaseous material, $DGR$ the dust-to-gas ratio, and $q_{\rm PAH}$ is the abundance of PAH molecules relative to the overall dust mass \citep[][and see discussion in \citealt{Leroy2023}, \citealt{Sandstrom2023}, \citealt{Chown2025}]{Draine2007}. We lack constraints for $DGR$ and $q_{\rm PAH}$ in the outflow, but $N(H)$ and $U$ change with elevation \citep{Leroy2015}.

Figure~\ref{fig:profiles} explores this behavior. The top row shows the profiles of PAH intensity (here \textit{Spitzer} $8\mu$m\footnote{We use \textit{Spitzer} as it closely matches F770W and covers the full region, minimizing edge effects.}) along with surface densities of H$_2$ traced by CO(1--0), atomic gas (HI), and the total neutral gas surface density ($\mathrm{\Sigma_{HI}+\Sigma_{H_2}}$), as a function of distance along the minor axis. We have convolved all the data in this section to 27\arcsec{} and reproject onto a shared astrometric grid. The second row shows the PAH emission normalized by the molecular, atomic, and total gas column density as a function of distance from the starburst. The final row shows three estimates of the intensity of the radiation field: $G/G_0$ from PDR modeling \citep{Levy2023}, $\langle U \rangle$ from IR SED modeling \citep{Leroy2015} (with models from \citealt{Draine2007}), and one assuming $U \propto 1/r^2$, where $r$ is the radial distance from the galaxy center. Following Eq. \ref{eq:ipah}, we expect $I_{\rm PAH} / \Sigma_{\rm HI+H_2} \propto U$.

From the top panels, gas surface densities and PAH emission all decline with distance. The PAHs and H$_2$ fall similarly, and the HI decreases more shallowly than both of them. The HI shows a flat profile and exceeds the H$_2$ so that most neutral gas beyond $\gtrsim 1$~kpc is HI in the north and south (modulo some uncertainty in $\alpha_{\rm CO}$ indicated by the shaded region). From the middle panels, the ratios of PAH emission to gas surface densities fall rapidly near the disk but flatten significantly past 750~pc. The flat PAH-to-H$_2$ ratio is consistent with the nearly linear relation between CO and PAH emission in Figure~\ref{fig:whole_correlation} and the stable but noisy ratios in Figure \ref{fig:ratios}. Though HI makes up most of the neutral mass at high elevation, the PAH intensity to total gas ratio varies only weakly in both the north and the south.

In the bottom panels we compare the PAH/$\Sigma_{\mathrm{HI+H_2}}$ profile to estimates of the intensity of the interstellar radiation field. We consider the $\langle U \rangle$ from IR SED modeling the most reliable estimate of the intensity of radiation illuminating the PAHs. This shows a somewhat shallower dependence than $1/r^2$ over the range that we study, indicating that radiative transfer and opacity effects may complicate the radiation field beyond simple illumination by the starburst.

The mild decline in the PAH-gas ratio is consistent with the decline of the $\langle U \rangle$ profile over the same range. Thus the observations appear consistent with Equation~\ref{eq:ipah}, with the PAHs mixed with all neutral gas components and responding to heating by the interstellar radiation field. We do not have independent constraints on $q_{\rm PAH}$ and $DGR$\footnote{Our adopted $\alpha_{\rm CO}$ from \citet{Leroy2015} assumes a fixed $DGR$ over the range of our study.}. The simulations of \cite{Richie2025} find that $DGR$ decreases with distance from the starburst due to dust destruction. Any decrease in $DGR$ (or $q_{\rm PAH}$, e.g., due to selective PAH destruction) would decrease $I_{\rm PAH}/\Sigma_{HI+H2}$ at fixed $U$. Given the good match between $U$ and $I_{\rm PAH}/\Sigma_{\rm HI+H2}$ in the bottom panel of Figure \ref{fig:profiles}, such variations do not seem to be required by the data out to $\lesssim 2.5$~kpc.

\section{Conclusions}

We have compared PAH emission captured by JWST/MIRI and \textit{Spitzer} to the cold, warm, and hot gas traced by CO, H$\alpha$, and X-ray emission along the outflow in the prototypical starburst galaxy, M82. 

\begin{enumerate}

\item Within the outflow cone, which is the area mapped by JWST, all three gas tracers correlate with PAH emission (Figure~\ref{fig:whole_correlation}). On larger scales, currently only visible in \textit{Spitzer} PAH mapping, PAH emission show a much wider distribution than the X-ray and H$\alpha$ emission, which are confined to the outflow cone (Figure~\ref{fig:data_image}).

\item The CO and H$\alpha$ emission both show a moderate-to-strong correlation with PAH emission over the extent of the outflow covered by JWST. The X-rays, meanwhile decouple at distances $>2$~kpc. The correlation with X-rays may simply reflect a shared decline in intensity with height (Figure~\ref{fig:residuals}).

\item The close correspondence between CO and PAH emission likely reflects that PAHs are mixed with cold gas. As shown in Appendix~\ref{app:ha_zoomin}, the H$\alpha$ emission traces the surfaces of cold clouds, explaining the excellent correlation between the two in Figure~\ref{fig:whole_correlation} and Figure~\ref{fig:ratios}.

\item Surprisingly, the CO-to-PAH scaling relations that we find for both the F770W and F1130W filters closely resemble that in star-forming galaxy disks, despite the distinct conditions in the M82 outflow. This may reflect that regardless of environment, PAHs are excellent cold gas tracers.

\item In lower resolution vertical profiles, the ratio of PAH emission to total neutral (HI+H$_2$) gas falls slowly with increasing elevation over the range $0.75-2.5$~kpc (Figure~\ref{fig:profiles}). This decline matches variations in the intensity of the interstellar radiation field inferred from IR SED modeling. This appears to imply little variation in the product of PAH abundance and the dust-to-gas ($q_{\rm PAH} \times DGR$) over this part of the outflow.

\end{enumerate}

Future work will explore in depth how PAHs evolve in M82's outflow including: how do the F1130W-to-F770W ratio (tracing PAH charge) and the PAH-to-TIR ratio (tracing PAH abundance) change with distance along the outflow (Cronin et al.\ in prep.)? In addition, it would be useful to consider how other multiwavelength data like X-rays and H$\alpha$ compare to the 3.3~$\mu$m PAH feature captured by the F335M JWST filter, highlighting the evolution of the smallest, most fragile PAHs (Arens et al.\ in prep).

\facilities{JWST, HST, \textit{Chandra}, IRAM, \textit{Spitzer}}
\software{astropy \citep{astropy,Astropy2018,Astropy2022}, scipy \citep{2020SciPy-NMeth}}

\begin{acknowledgements}

This work is based on observations made with the NASA/ESA/CSA JWST.The data were obtained from the Mikulski Archive for Space Telescopes at the Space Telescope Science Institute, which is operated by the Association of Universities for Research in Astronomy, Inc., under NASA contract NAS 5-03127 for JWST. These observations are associated with program 1701.

S.L. and L.A.L. were supported by NASA’s Astrophysics Data Analysis Program under grant No. 80NSSC22K0496SL, and L.A.L. also acknowledges support through the Heising-Simons Foundation grant 2022-3533. 

A.K.L. and C.R. gratefully acknowledge support from NSF AST AWD 2205628, JWST-GO-01701.016-A. A.K.L. also gratefully acknowledges support by a Humbolt Research Award.

R.H.-C.\ thanks the Max Planck Society for support under the Partner Group project "The Baryon Cycle in Galaxies" between the Max Planck for Extraterrestrial Physics and the Universidad de Concepción. R.H-C. also gratefully acknowledge financial support from ANID - MILENIO - NCN2024\_112 and ANID BASAL FB210003. 
R.S.K.\ acknowledges financial support from the ERC via Synergy Grant ``ECOGAL'' (project ID 855130),  from the German Excellence Strategy via the Heidelberg Cluster ``STRUCTURES'' (EXC 2181 - 390900948), and from the German Ministry for Economic Affairs and Climate Action in project ``MAINN'' (funding ID 50OO2206).  RSK also thanks the 2024/25 Class of Radcliffe Fellows for highly interesting and stimulating discussions. 
L.A.B. acknowledges support from the Dutch Research Council (NWO) under grant VI.Veni.242.055 (\url{https://doi.org/10.61686/LAJVP77714}) and the ERC Consolidator grant 101088676 ("VOYAJ”).

\end{acknowledgements}

\bibliography{main}{}
\bibliographystyle{aasjournal}

\appendix
\setcounter{figure}{0}
\renewcommand{\thefigure}{A\arabic{figure}}
\section{The Effect of Vertical Distance Scaling on PAH-Gas Phase Correlations}
\label{sec:append_distance}
\begin{figure*}[h!]
    \centering
    \includegraphics[width=\textwidth]{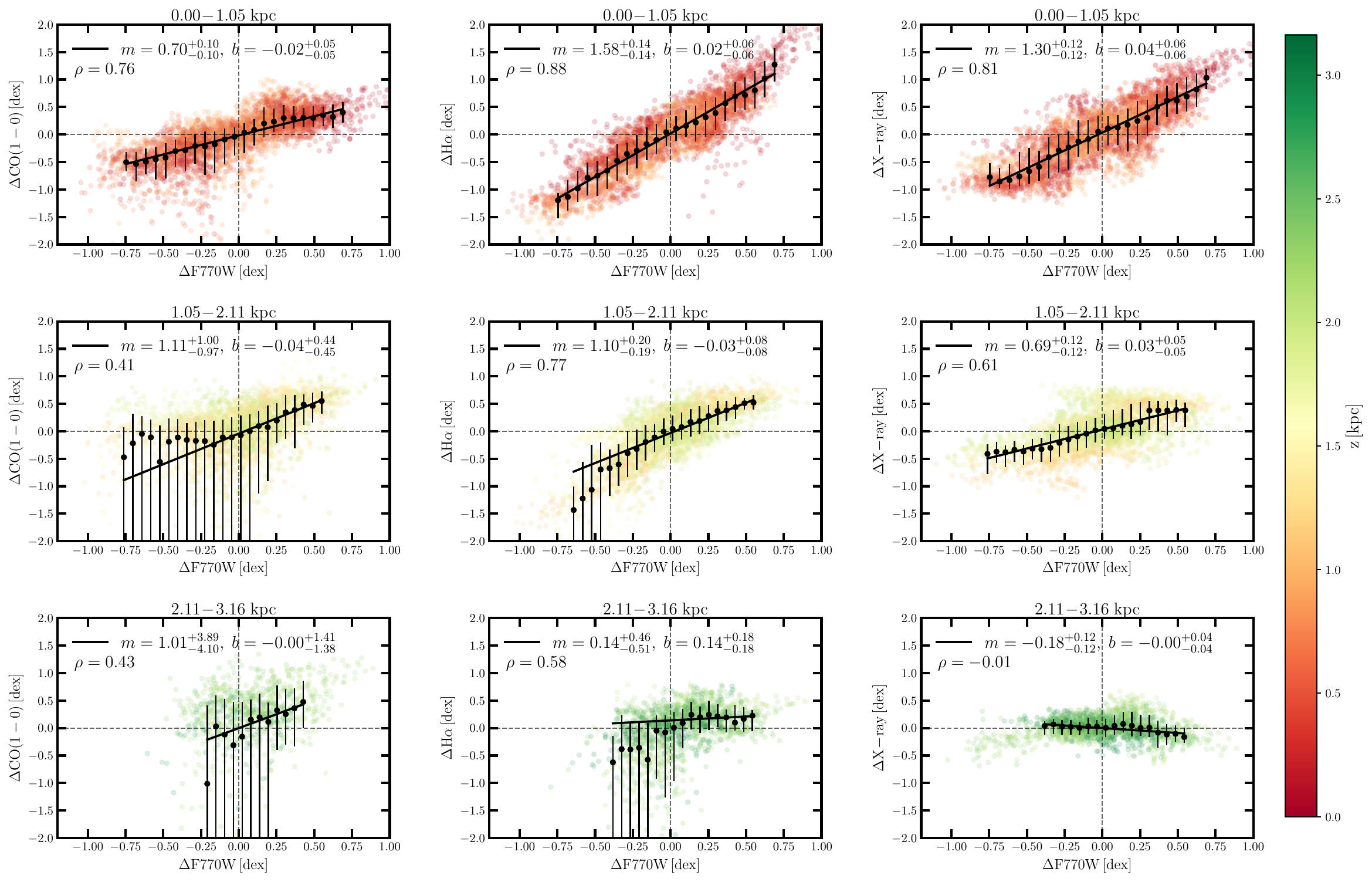}
    \caption{Residual plots showing gas phase intensity (y-axis) and F770W intensity (x-axis) corrected for distance. From left to right the columns are for CO(1--0), H$\alpha$, and X-ray. From top to bottom the rows are distance separations at 0--1.05~kpc, 1.05--2.11~kpc, and 2.11--3.16~kpc. The upper distance bound is the maximum distance from all datasets. The black points are the medians for the distribution of data points, and the error bars are the 16th and 84th percentiles. In the top left of each panel, we show the best power-law fits to the medians and Spearman rank correlation coefficient of the whole distribution. }
    \label{fig:residuals}
\end{figure*}

In Figure~\ref{fig:residuals} we analyze how the PAH-gas phase relationships scale when distance is fixed. To accomplish this for each wavelength, X-ray, H$\alpha$, F770W, and CO(1--0), we plot their intensity versus vertical height ($z$). We then create medians for these intensity-distance plots. For each dataset, and for each corresponding pixel, we subtract the median intensity value at the $z$ position of the original data pixel. The various phases are then plotted against the F770W data in Figure~\ref{fig:residuals}

We find that when correcting for vertical distance (as discussed in Section~\ref{ssec:pah_correlations}), only the CO(1--0) and H$\alpha$ datasets still show a moderate correlation with the F770W data past 2~kpc. The X-ray data does not present any correlation (Spearman rank coefficient of near zero), showing that the dust and hot gas are decoupled at these large distances and only correlate near the starburst. We find near identical results for the F1130W (not shown).

\setcounter{figure}{0}
\renewcommand{\thefigure}{B\arabic{figure}}
\section{H$\alpha$ Interface Zoom In}
\label{app:ha_zoomin}

\begin{figure*}
    \centering
    \includegraphics[width=\textwidth]{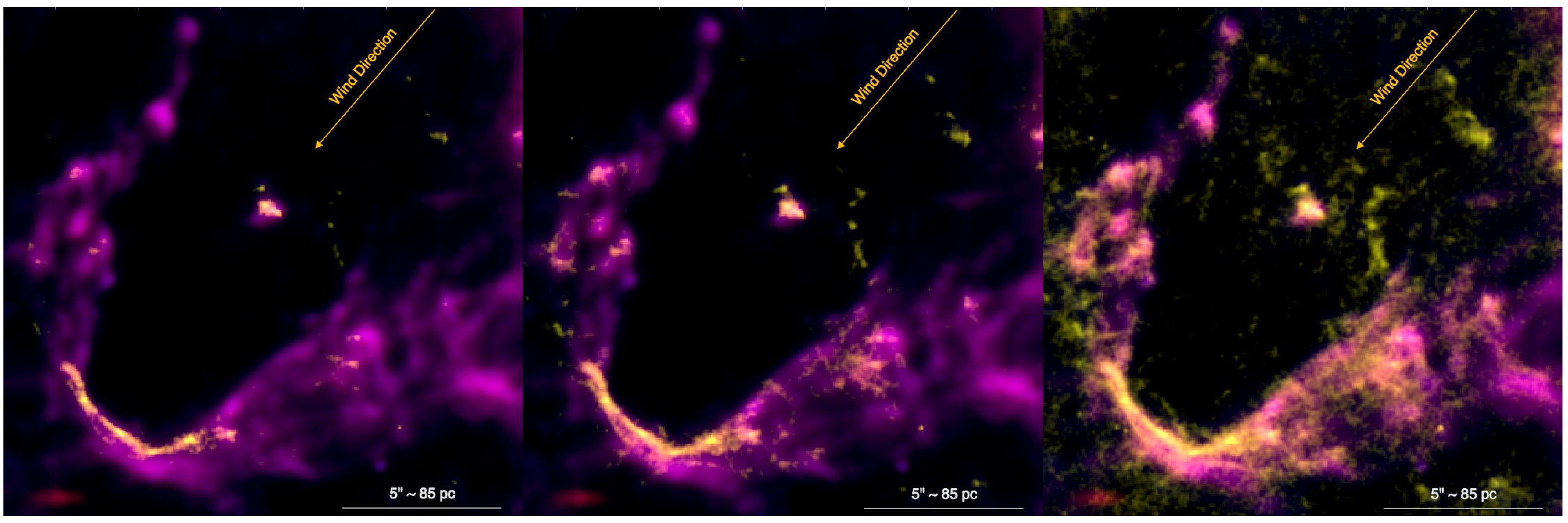}
    \caption{Three color image of F770W in blue, F1130W in red, and H$\alpha$ in yellow. The H$\alpha$ appears at the surface of the PAH cloud that is tracing colder gas interacting with the hot wind. From left to right we increase the stretch of the image to show how the H$\alpha$ envelopes the cloud, tracing the mixing layer between cloud and wind. }
    \label{fig:ha_cloud}
\end{figure*}

In Figure~\ref{fig:ha_cloud}, we present a zoom-in on cloud 12 from \citet{Lopez2025}, which exhibits an arc-like structure curving away from the wind direction. Strikingly, the same feature is visible in our JWST PAH images (F770W and F1130W). The H$\alpha$ emission lies along the outer layers of the cloud, preferentially on the wind-facing side, as indicated by the vector in Figure~\ref{fig:ha_cloud}.

As hypothesized by \citet{Lopez2025}, the H$\alpha$ emission may trace turbulent mixing layers modeled in multiphase galactic wind simulations \citep{Tan2024}, which can cool rapidly and prolong cloud survival. As shown in the main text, the strong correlation between molecular gas and PAH emission demonstrates that JWST images provide a high-resolution proxy for cold gas. Thus, Figure~\ref{fig:ha_cloud} likely captures the interfaces between PAH-traced cold clouds and the hot wind, where H$\alpha$ emission arises from their surfaces, presumably due to shocks.

\end{document}